\documentclass[[aps,preprintnumbers,prd,twocolumn,superscriptaddress]{revtex4}

\usepackage{amsfonts,amssymb,amsmath,color,xcolor}
\usepackage{float}
\usepackage{graphicx}
\graphicspath{ {./images/} }
\usepackage[colorlinks=true,linkcolor=red,urlcolor=red,citecolor=red]{hyperref}
\begin{document}

\title{Can wormholes and black holes be distinguished by magnification?}
\author{Ke Gao}
\email{2021700389@stu.jsu.edu.cn}
\author{Lei-Hua Liu}
\email{liuleihua8899@hotmail.com}

\affiliation{Department of Physics, College of Physics, Mechanical and Electrical Engineering, Jishou University, Jishou 416000, China}

\begin{abstract}
The magnification effect of wormholes and black holes has been extensively researched. It is crucial to provide a finite distance analysis to understand this magnification phenomenon better. In our article, the rotational Simpson-Visser metric (RSV) is chosen as the focus of research. By calculating the deflection angle of light in RSV metric, we determine the resulting magnification effect, then applied the RSV metric to specific examples such as an Ellis-Bronnikov wormhole, a Schwarzschild black hole, and a Kerr black hole (or wormhole) to analyze the magnification. We find that an Ellis-Bronnikov wormhole has only a single peak of magnification, while a Kerr black hole has one to three magnification peaks. In addition, the article's findings suggest that the lensing effect of the Central Black Hole of the Milky Way Galaxy exhibits multiple peaks of magnification. Our research provides the possibility of distinguishing between wormholes and black holes from a phenomenological perspective.
\end{abstract}

\maketitle
\bigskip

\section{Introduction}
\label{intro}
\begin{figure}
    \centering
    \includegraphics[scale=1]{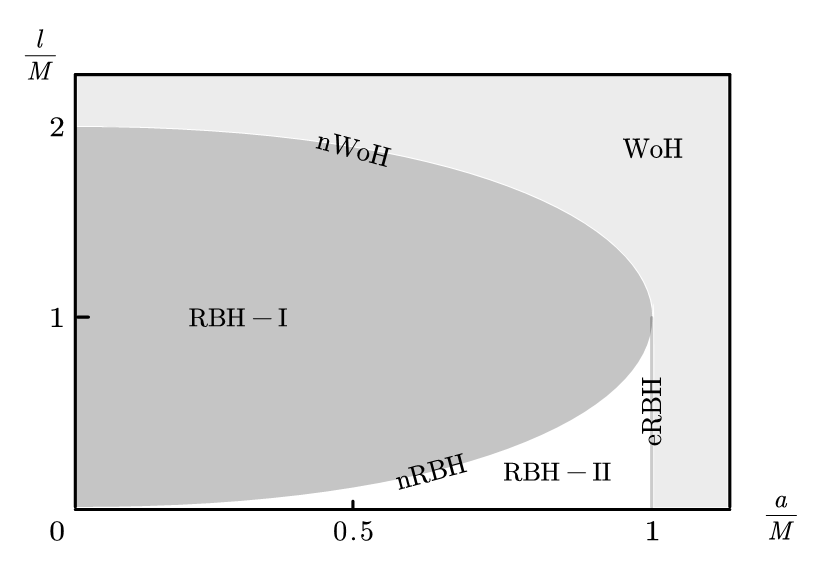}
    \caption{Parameter space and corresponding spacetime structure of RSV. WoH indicates traversable wormhole;
nWoH denotes null WoH, i.e. one-way wormhole with the null throat;
RBH-I expresses regular black hole with one horizon (in the $r>0$ side, plus its mirror image in the
$r<0$ side);
RBH-II signifies regular black hole with an outer and an inner horizon (per side);
eRBH stand for extremal regular black hole (one extremal horizon per side);
nRBH null RBH-I, i.e. a regular black hole with one horizon (per side) and a null throat.
(Image referenced from \cite{Mazza:2021rgq}). }
    \label{fig: 1}
\end{figure}

The recent development of gravitational wave astronomy has brought a shift in the discussion surrounding black hole mimickers from theory to empirical observations \cite{Mazza:2021rgq, Annulli:2021ccn, Liu:2021aqh, Bao:2018msr, Abdulxamidov:2022ofi, Heckman:2021vzx, Aneesh:2018hlp, Wang:2018mlp, Simpson:2019cer, Bambi:2021qfo}. With the absence of a definitive quantum gravity theory \cite{Clifton:2011jh, Capozziello:2011et}, there is an attraction to explore alternative phenomenologically viable scenarios using simple meta-geometries. This approach offers a compact and accessible means of investigating such possibilities.  Simpson and Visser \cite{Mazza:2021rgq, Simpson:2019cer, Lobo:2020ffi} have put forward a static and spherically symmetric metric known as the SV metric, which provides a comprehensive and nuanced model for black holes and wormholes. This metric smoothly transitions between different cases and is described by the line element:

\begin{equation}
\begin{split}
    ds^2&=-\left(1-\frac{2M}{\sqrt{r^2+l^2}}\right)dt^2+\left(1-\frac{2M}{\sqrt{r^2+l^2}}\right)^{-1}dr^2\\
   &+\left(r^2+l^2\right)\left(d\theta^2+\sin^2\theta d\phi^2\right),
\end{split}
\end{equation}
where $M\geqslant 0$ represents the ADM mass and $l \geqslant 0$ is a parameter responsible for regularizing the central singularity (when $M=0$, $l$ is the throat of Ellis-Bronnikov wormholes). By employing the Newman–Janis procedure \cite{Newman:1965tw}, the rotating SV metric (RSV) can be obtained:
\begin{equation}
\label{RSV}
\begin{split}
    ds^2&=-\left(1-\frac{2M\sqrt{r^2+l^2}}{\Sigma}\right)dt^2+\frac{\Sigma}{\Delta}dr^2\\
   &+\Sigma d\theta^2-\frac{4Ma\sin^2\theta\sqrt{r^2+l^2}}{\Sigma}dtd\phi+\frac{\chi\sin^2\theta}{\Sigma}d\phi^2,
\end{split}
\end{equation}
where $\Sigma=r^2+l^2+a^2\cos^2(\theta)$, $\Delta=r^2+l^2+a^2-2M\sqrt{r^2+l^2}$, and $\chi=(r^2+l^2+a^2)^2-\Delta a^2\sin^2\theta$. Here, $a$ represents the ratio of spin angular momentum $J$ to mass $M$, $a=\frac{J}{M}$. The RSV metric \eqref{RSV} transforms to the SV metric when $a=0$ and to the Kerr metric when $l=0$. The values of $\frac{l}{M}$ and $\frac{a}{M}$ determine the conversion between black holes and wormholes in the RSV metric, as shown in FIG. \ref{fig: 1}. These distinctive features of RSV make it a convenient tool for studying the differences between black holes and wormholes.

Gravitational lensing is a powerful technology for exploring the universe \cite{Huang:2023bto,Kelly:2023mgv,Li:2023zrz,Awad:2023tvu}. The presence of a lens can alter the space-time and impact the observed physical quantities such as magnification \cite{Kelly:2023wzm,Awad:2023tvu,Gao:2022cds,Sun:2022ujt}, event rate \cite{Kelly:2023mgv,Cai:2022kbp,Gao:2023sla} and shadow \cite{Meng:2023wgi,Ramadhan:2023ogm,Xavier:2023exm,Azreg-Ainou:2014dwa}. By studying these observable measurements, we can explore the lens itself. Before our work, physicists have extensively studied the lensing effect of black holes \cite{Delos:2023fpm,Lu:2022yuc,Sun:2022ujt,Verma:2022pym,Gao:2022cds,Rose:2022xdu} and wormholes \cite{Cheng:2021hoc,Tsukamoto:2017hva,Sajadi:2016hko,Liu:2022lfb,Tsukamoto:2016zdu,Safonova:2001vz,Torres:1998xd,Nascimento:2020ime,Tsukamoto:2020bjm,Islam:2021ful}. Their research provides valuable insights that inspire us to search the possibility of distinguishing wormholes and black holes based on magnification: wormholes and black holes may exhibit different magnification patterns.

In this article, we select the RSV metric as the research object because it can smoothly transform between wormholes and black holes. We calculate the deflection angle of light in the RSV metric and study the resulting magnification effects. We specifically discuss the magnification effects of Ellis-Bronnikov wormhole, Schwarzschild black hole,  and Kerr black hole, where we use the dimensionless units to analyze the variation of magnification with lens geometry, wormhole throat radius, ADM mass, and spin. At the end of the article, we restore physical units to analyze the magnification of the Central Black Hole of the Milky Way Galaxy.

The structure of this article is as follows: the first section \ref{intro} provides an introduction to background space and gravitational lensing. Section \ref{spacetime} describes some geometric quantities in axisymmetric rotational spacetime. Section \ref{deflection angle} calculates the deflection angle of the RSV metric. Section \ref{magnification} analyzes the magnification effect of RSV associating spacetime. Section \ref{conclusions} concludes and provides future prospects.

\section{Spacetime}
\label{spacetime}
In this section, we will explore various geometric quantities associated with the metric \eqref{RSV}. We use the terminology following the literature references \cite{Ono:2017pie, Ono:2018jrv, Ishihara:2016sfv, Ishihara:2016vdc}.
For convenience, let us rewrite the metric \eqref{RSV} as:
\begin{equation}\label{eq 3}
ds^2=-Adt^2 + Bdr^2 + Cd\theta^2 - 2Hdtd\phi + Dd\phi^2,
\end{equation}
where $A=g_{tt}$ in the metric \eqref{RSV}, $B=g_{rr}$, and so on.
Our primary focus is on investigating the motion of photons on the equatorial plane with $\theta=\frac{\pi}{2}$ in the metric \eqref{eq 3}. The length along the path of light, denoted as $ds^2=0$, can be expressed as:
\begin{equation}
d\ell^2\equiv \gamma_{ij}dx^idx^j=\frac{B}{A}dr^2+ \frac{C}{A}d\theta^2+\frac{AD+H^2}{A^2}d\phi^2.
\end{equation}
Here, $\gamma_{ij}$ defines a 3-dimensional Riemannian space where the photon's motion is described as a trajectory in a spatial curve $\ell$.
The closest distance between a photon and the central celestial body in the lens plane is known as the impact parameter $b$ (the definition comes from straight-line approximation), which can be expressed as:
\begin{equation}
b\equiv \frac{L}{E}=\frac{-H+D\frac{d\phi}{dt}}{A+H\frac{d\phi}{dt}}.
\end{equation}
This definition above is general for axisymmetric spacetime.
$L$ represents the angular momentum of the photon and $E$ represents its energy.
In terms of the impact parameter, the orbital equation of the photon can be expressed as:
\begin{equation}
\left(\frac{dr}{d\phi}\right)^2=\frac{AD+H^2}{B}\frac{D-2Hb-Ab^2}{(H+Ab)^2}.
\label{eq 6}
\end{equation}
This equation can be solved to obtain $r=F(\phi)$ using perturbational methods (see the appendix of \cite{Gao:2023ltr}). At a zero-order approximation, we can obtain $r=\frac{b}{\sin\phi}$. In metric \eqref{eq 3}, the geodesic curvature can be defined as:
\begin{equation}
\kappa=-\frac{1}{\sqrt{\gamma\gamma^{\theta\theta}}}\partial_r\left(\frac{H}{A}\right),
\end{equation}
and the Gaussian curvature as:
\begin{small}
\begin{equation}
K=-\sqrt{\frac{A^3}{B(AD+H^2)}}\partial_r\left[\frac{1}{2}\sqrt{\frac{A^3}{B(AD+H^2)}}\partial_r\left(\frac{AD+H^2}{A^2}\right)\right].
\end{equation}
\end{small}
With these preparations, we can now investigate the lensing effect of the RSV metric.

\begin{figure}
    \centering
    \includegraphics[scale=0.6]{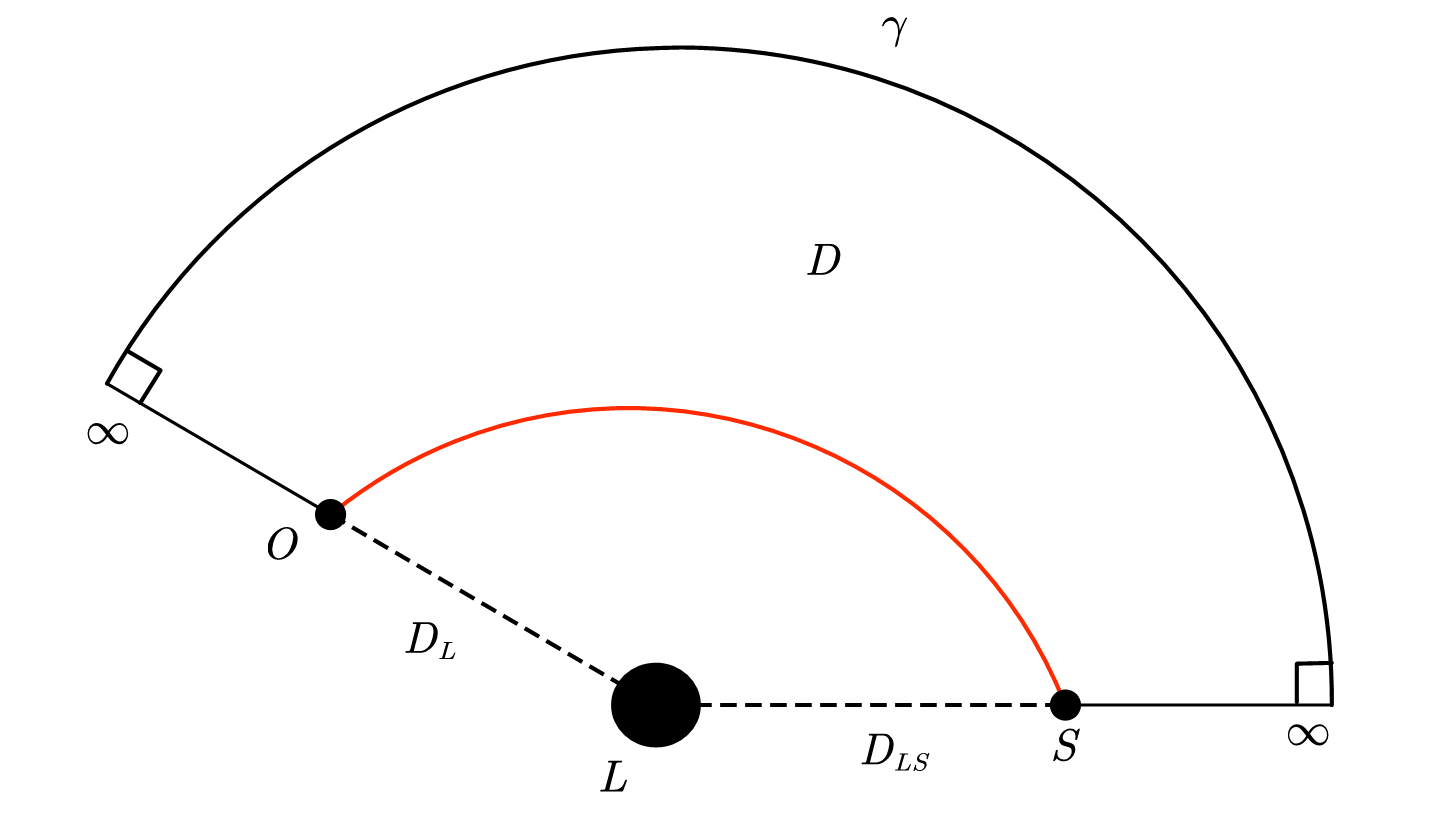}
    \caption{Displaying the Gaussian Bonnet Integral Domain. The spin of the celestial body will make the original geodesic no longer be the geodesic. }
    \label{fig:1}
\end{figure}

\begin{figure}
    \centering
    \includegraphics[scale=0.8]{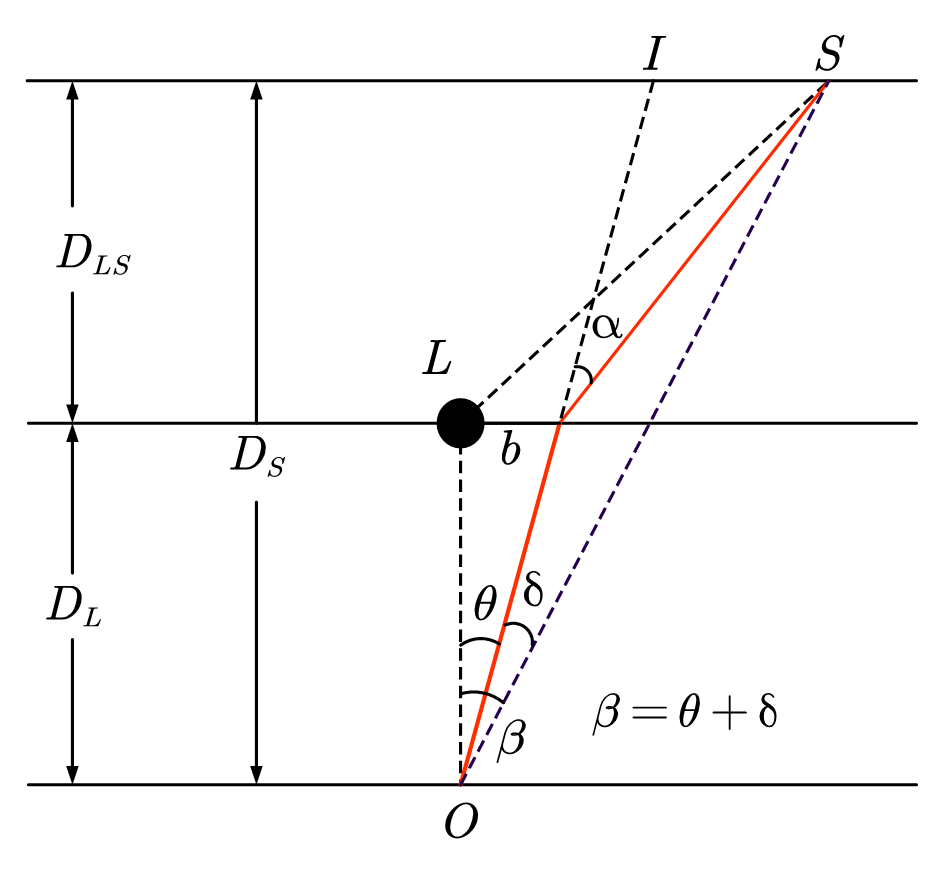}
    \caption{Showing the lens plane, $L$ represents wormholes or black holes, O is the observer, S is the source, $D_{LS},D_L,$ and $ D_S$ is the angular radius distance. $b$ is the impact parameter, while others are angular quantities.}
    \label{fig:2}
\end{figure}

\section{deflection angle}
\label{deflection angle}
We consider that both the source and the observer are far away from the lens.
The deflection angle $\alpha$ is defined as the difference between the ray directions at the
source $\ell=-\infty$ and the observer $\ell=\infty$ in the asymptotically flat regions \cite{Asada:2003nf},
\begin{equation}
\alpha_{\text{infinity}}=e_S-e_O=-\int_{-\infty}^\infty d\ell \frac{de}{d\ell}.
\end{equation}
 We adopt GW method \cite{Gibbons:2008rj, Werner:2012rc} to calculate the deflection angle by Gaussian Bonnet theory:
\begin{equation}
\int\int_D KdS+\int_{\partial D}\kappa d\ell+\sum\limits_i\alpha_i=2\pi\chi(D).
\end{equation}
One can choose the integral domain $D$ in Fig. \ref{fig:1}, $SO$ is the path of light. Besides, this Euler index $\chi$ in domain $D$ is one. Then,
\begin{equation}
\int\int_{D}KdS+\int_{\gamma}\kappa d\ell+\int_{OS}\kappa d\ell+\sum_i\alpha_i=2\pi.
\end{equation}
We can set $\gamma$ to vertically intersect extension line of $SO$ at infinity, which means
\begin{equation}
\sum_i\alpha_i=\frac{\pi}{2}(\infty)+\frac{\pi}{2}(\infty)+\rho_s+\rho_o=\pi+\rho.
\end{equation}
Here, $\rho=\rho_s+\rho_o$ is the sum of external angles at points O and S.
Then, one do an integral transformation
\begin{equation}
\kappa d\ell=\kappa\frac{d\ell}{d\phi}d\phi.
\end{equation}
Here, $\phi$ is the angular coordinate of the center at $L$. We only fixed the endpoint of $\gamma$, so we can directly set up $\kappa\frac{dt}{d\phi}=1$ on $\gamma$, which leads to
\begin{equation}
\int\int_{D}KdS+\int_{0}^{\pi-\rho+\alpha_{\text{finite}}}d\phi+\int_{OS}\kappa d\ell+\pi+\rho=2\pi.
\end{equation}
 Here, $\angle(LS,LO)+\rho_s+\rho_o+(\pi-\alpha_{\text{finite}})=2\pi$, we select the $\phi$ coordinate of point O to be zero. Now we assume that both the observer and the light source are far away from the lens center which means $D_L\to\infty$ and $D_{LS}\to\infty$. Then $\alpha_{\text{finite}}\to \alpha_{\text{infinite}}$,
\begin{equation}
\label{eq 14}
\alpha_{\text{infinite}}=-\int_{\phi_O}^{\phi_S}\int_r^\infty KdS+\int_{\phi_O}^{\phi_S}\kappa\psi d\phi,
\end{equation}
where used $d\ell=\psi d\phi$. 
In the case of metric \eqref{RSV}, we present the results of each part of the calculation:
\begin{equation}
\label{eq.r}
\frac{1}{r}=\frac{\sin\phi}{b}+\frac{M}{b^2}(1+\cos^2\phi)+\mathcal{O}(M^2,a^2,l^2,Ma),
\end{equation}
\begin{equation}
K=-\frac{2M}{r^3}+\frac{3M^2}{r^4}-\frac{l^2}{r^4}+\mathcal{O}(l^3,a^3,M^3,...),
\end{equation}
\begin{equation}
dS=\left(r+\frac{l^2}{2r}+3M+\frac{15M^2}{2r}+\mathcal{O}(l^3,a^3,M^3,...)\right)drd\phi,
\end{equation}
\begin{equation}
\kappa=\pm\frac{2aM}{r^3}+\mathcal{O}(l^3,a^3,M^3,...),
\end{equation}
and
\begin{equation}
\psi= b\csc^2(\phi)+\mathcal{O}(l,M,a).
\end{equation}
Ultimately, the infinite distance deflection angle of the second order with weak field approximation is obtained as
\begin{equation}
\label{eq.15}
\alpha_{\text{infinite}}=\frac{4M}{b}+\frac{\pi  \left(l^2+15 M^2\right)}{4 b^2}\pm\frac{4aM}{b^2},
\end{equation}
where $\pm$ comes from the spin direction.
 For the deflection angle $\alpha_{\text{infinite}}$, when $M$ is zero, it is regained to the deflection angle of the Ellis-Bronnikov wormhole \cite{Dey:2008kn} where $l$ is the wormhole throat. When $l^2$ equals to zero, it is recovered to the Kerr black hole case \cite{Werner:2012rc}. Referring to Fig. \ref{fig:1} and \ref{fig:2}, we obtain a finite distance deflection angle:
 \begin{equation}
\alpha_{\text{finite}}=\bigg[\frac{l^2+15M^2}{4b^2}\phi-\frac{2M}{b}\cos\phi\pm \frac{2aM}{b^2} \cos\phi \bigg]_{\phi=\phi_O}^{\phi=\phi_S}
 \end{equation}
with
\begin{equation}
\label{eq 23}
\phi\big|_{\phi_O}^{\phi_S}=\pi-\arcsin\frac{b}{D_{LS}}-\arcsin\frac{b}{D_L}
\end{equation}
and
\begin{equation}
\label{eq 24}
\cos\phi \big|_{\phi_O}^{\phi_S}=-\big(\sqrt{1-b^2/D_{LS}^2}+\sqrt{1-b^2/D_L^2} \big).
\end{equation}
These geometric relationships can also be obtained by solving Eq. \eqref{eq 6}, and then the angular coordinates $\phi_S$ and $\phi_O$ correspond to the distances $D_{LS}$ and $D_L$,  respectively. Furthermore, we neglect the term of $\mathcal{O}(M)$ in Eq. \eqref{eq 23} and Eq. \eqref{eq 24}. For more details, see \cite{Ono:2017pie}.

\section{magnification}
\label{magnification}

Magnification reflects the degree to which light is distorted, which is an observable quantity. We will study the lensing effect from this perspective. Referring to Fig. \ref{fig:2}, one can obtain the geometric relationship as
\begin{equation}
\label{eq.16}
\beta=\theta-\frac{D_{LS}}{D_S}\alpha.
\end{equation}
The lens potential $\Psi$ which can connect deflection angle and magnification is defined as
$
\Psi\equiv\frac{2D_{LS}}{D_LD_S}\int\Phi(\theta D_L,x)dx,
$
where $\Phi$ is Newtonian potential. The relationship between lens potential and deflection angle is as follows
\begin{equation}
\partial_\theta\Psi=D_L\partial_b\frac{2D_{LS}}{D_LD_S}\int\Phi(b,x)dx=\alpha.
\end{equation}
On the other hand, the Jacobian metric is also related to the lens potential
\begin{equation}
A_{ij}\equiv\left(\delta_{ij}-\frac{\partial^2\Psi}{\partial\theta_i\partial\theta_j}\right).
\end{equation}
The magnification is defined as the inverse of the Jacobian determinant and can, in axisymmetric spaces, be expressed as:
\begin{equation}
\label{eq.20}
\mu\equiv\frac{1}{\det A_{ij}}=\bigg|\frac{\beta}{\theta}\frac{d\beta}{d\theta}\bigg|^{-1}.
\end{equation}
Based on Eqs. \eqref{eq.20}, \eqref{eq.15} and \eqref{eq.16}, one could work out the  magnification of light in metric \eqref{RSV}:
\begin{equation}
\mu_{\text{finite}}=\bigg|\frac{D_L^2}{b}\beta \frac{\partial}{\partial b}\bigg(\frac{b}{D_L}-\frac{D_{LS}}{D_S}\alpha_{\text{finite}} \bigg) \bigg|^{-1}.
\end{equation}
Rather than discussing the analytical expression of magnification in which $\frac{d\mu}{db}$ and $ \frac{d^2\mu}{db^2}$ are characteristic curves, we focus on presenting numerical graphs of the magnification effect at finite distances.
\begin{figure}
    \centering
    \includegraphics[scale=1]{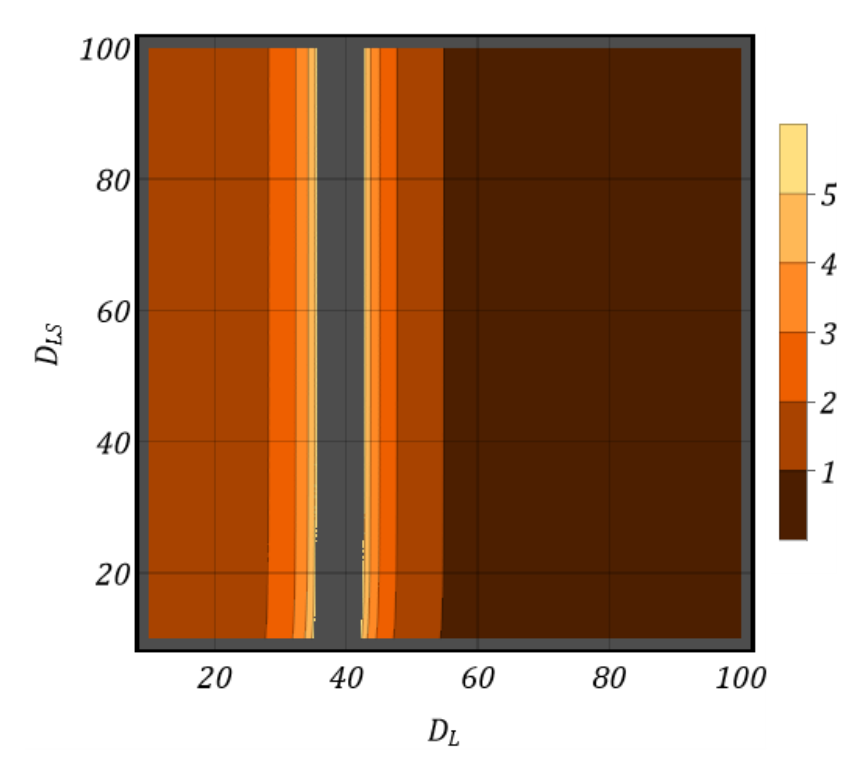}
    \caption{Contour plot of magnification of RSV. The relevant parameters are fixed as $M=0.03,\,\,l=0.05,\,\,a=0.1,\,\,b=2$. The black bar in the figure represents the position where the magnification peak appears.}
    \label{fig:4}
\end{figure}
\begin{figure}
    \centering
    \includegraphics[scale=1]{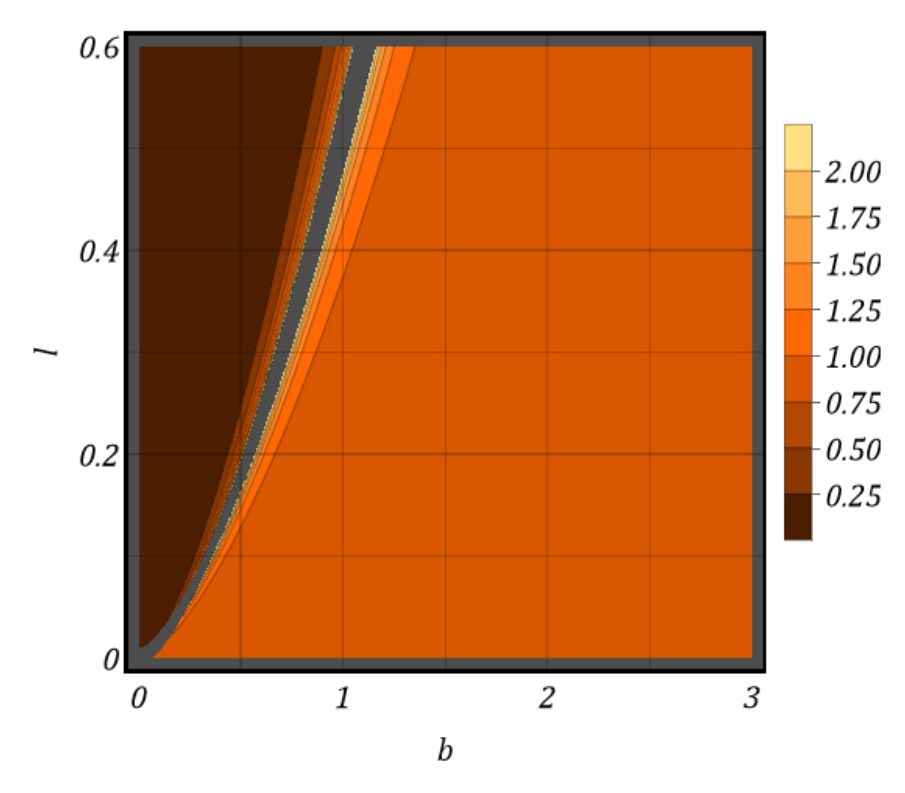}
    \caption{Contour plot of magnification of Ellis-Bronnikov wormhole.}
    \label{fig:5}
\end{figure}

We first consider the effect of lens geometry on magnification. (Due to the dimensionless nature of magnification, we adopt a dimensionless specification.)
The numerical results presented in Fig. \ref{fig:4} demonstrate that the distance between the source and lens has minimal influence on magnification, while the distance between the observer and lens has a noteworthy impact. As $D_L$ increases, the magnification decreases to one in the interval $D_L\in [55,100]$. and the magnification reduces to zero in the interval $D_L\in [100,\infty)$. In light of these findings, we will continue our discussion using a fixed parameter of $D_L=D_{LS}=10$ in subsequent cases.

If the ADM mass is set to zero, the RSV metric will transform into an Ellis-Bronnikov wormhole. In this scenario, where $l$ represents the throat radius of the wormhole, the magnification is depicted in Fig. \ref{fig:5}. From the figure, we can observe that the Ellis-Bronnikov wormhole exhibits a single peak of magnification. (How to determine the number of peaks? We use a straight line that is parallel to the b-axis to capture the black bars in the graph, and the number of cutoff points is the number of peaks of magnification.)
\begin{figure}
    \centering
    \includegraphics[scale=1]{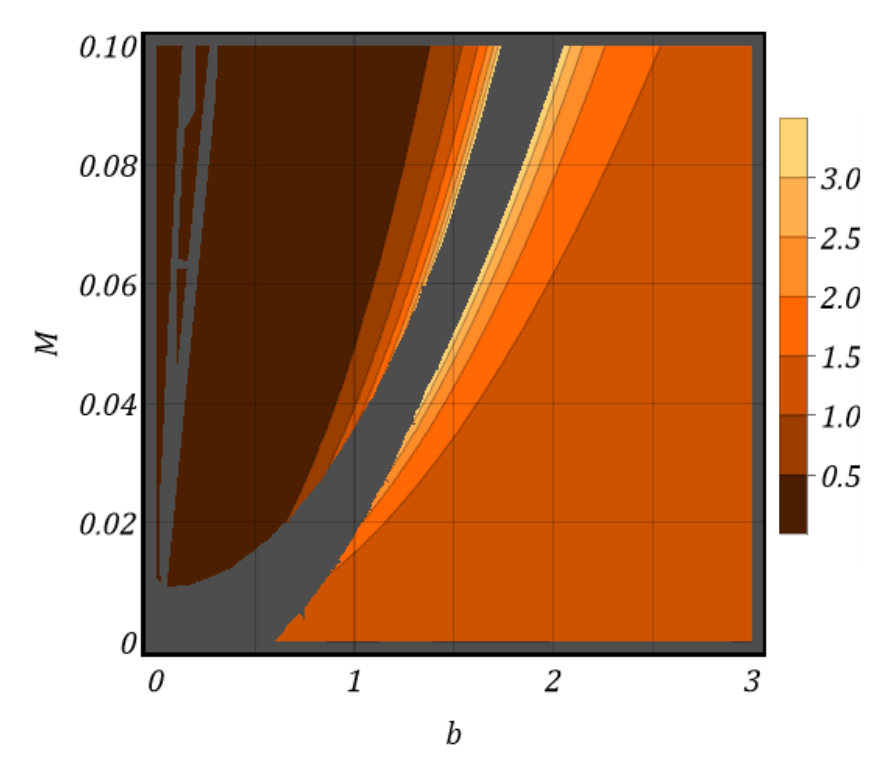}
    \caption{Contour plot of magnification of Schwarzschild black hole.}
    \label{fig:6}
\end{figure}

When both $l$ and $a$ are set to zero, the RSV metric reverts to a Schwarzschild black hole. In Fig. \ref{fig:6}, it can be observed that there are three peaks of magnification within the interval $M\in[0.05,0.10] $. As we decrease the mass to the range of $M\in[0.05,0.01]$, the three peaks coalesce into two. Finally, when the mass is less than $0.01$, only one peak of magnification remains. 
\begin{figure}
    \centering
    \includegraphics[scale=1]{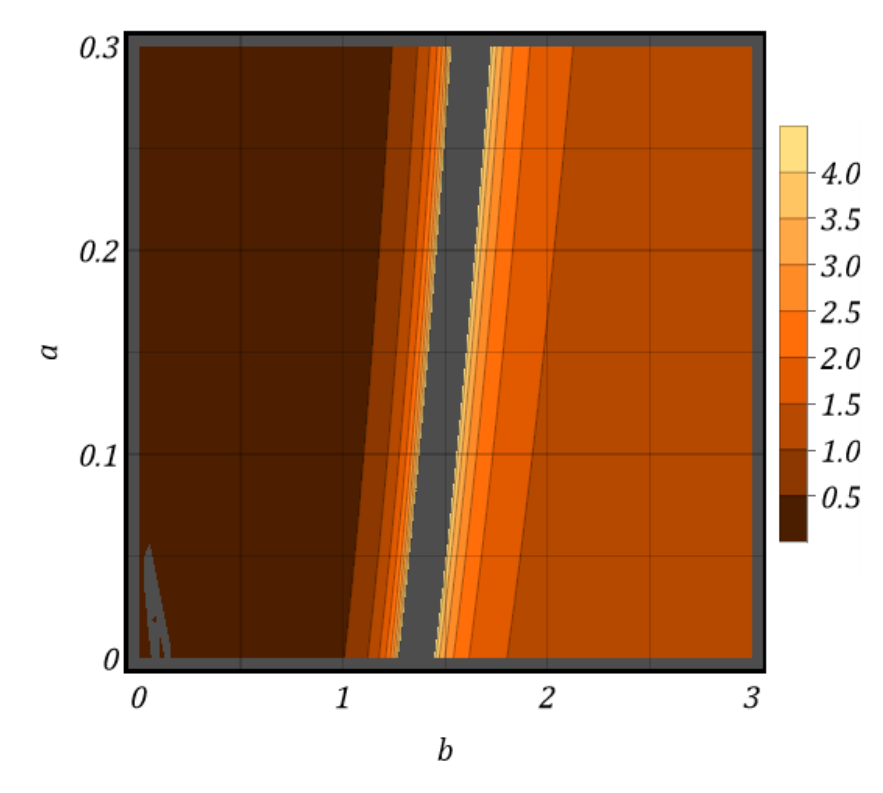}
    \caption{Kerr black hole with a positive spin, in which $M=0.05$. According to Fig. \ref{fig: 1}, when $a>M=0.05$, RSV will transform into a traversable wormhole.}
    \label{fig:7}
\end{figure}
\begin{figure}
    \centering
    \includegraphics[scale=1]{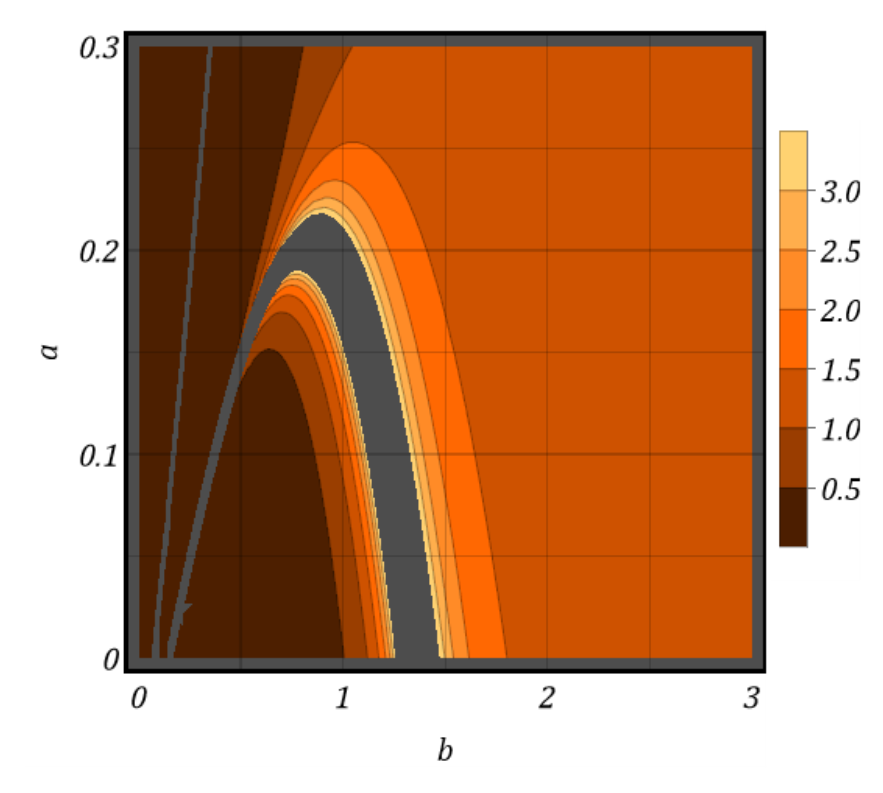}
    \caption{Kerr black hole with a negative spin, in which $M=0.05$.}
    \label{fig:8}
\end{figure}
In the case of a rotating black hole or possibly a wormhole, we set $l$ to zero.
It is significant to consider the direction of the black hole's spin determined by parameter $a$. We use the positive and negative cases of $a$ to represent different spin directions, as discussed in Fig. \ref{fig:7} and Fig. \ref{fig:8}, respectively.  
Let us first examine the scenario where the spin is positive, as depicted in Fig. \ref{fig:7}. In the interval of $a\in[0.00,0.05]$, two and three peaks of magnification can be observed. However, for $a\in[0.05,0.30]$, only one peak of magnification is present. Next, we consider the case of negative spin, as illustrated in Fig. \ref{fig:8}. Within the interval of $a\in, [0.00,0.18]$, three peaks of magnification exist. In the interval of $a\in[0.18,0.22]$, there are two peaks of magnification. Finally, for $a\in[0.22,0.30]$, only one peak of magnification is observed. 
It can be seen in Fig. \ref{fig:7} and \ref{fig:8} that $a$ equals to zero and corresponds to a Schwarzschild black hole with three peaks, while $a$ exists in the second-order term, which affects the position of the peaks and causes the two peaks to merge and finally disappear. 
\begin{figure}
    \centering
    \includegraphics[scale=0.75]{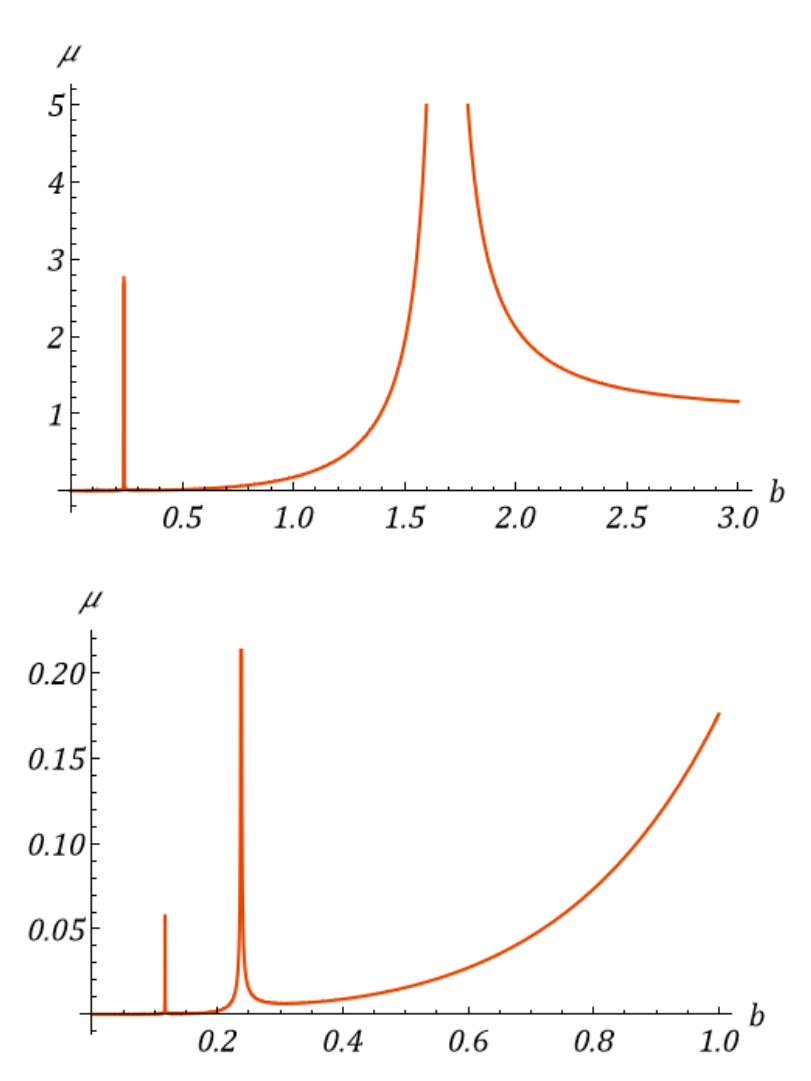}
    \caption{Schematic diagram of the three peaks of magnification curve of Schwarzschild black hole, corresponding to the case of $M=0.08$ in Fig. \ref{fig:6}. Here, the Schwarzschild radius about is $0.1$ on the $b$ axle.}
    \label{fig:9}
\end{figure}

\section{Conclusions and outlook}
\label{conclusions}
In this paper, we study the gravitational lensing effect of the RSV metric. 
We obtained the second-order deflection angle of the RSV metric and made finite distance corrections to it. Our analysis of magnification shows that the distance between the lens and the observer has a significant impact on the numerical value of magnification, which also indicates that finite distance correction is meaningful. On this basis, Ellis-Bronnikov Wormhole only exhibits a single peak of magnification, while Schwarzschild's black hole exhibits up to three peaks of magnification as the ADM mass increases. Black holes with negative spin return from three peaks to a single peak as the spin increases, and the same applies to the case of positive spin.

For the galaxy we live in, the mass of its central black hole is about four million solar mass $\rm M_\odot$. If we assume that its spin is small (Spin slightly affects the position of the peak) and be approximated as the Schwarzschild black hole situation, according to our calculations $\frac{\text{four million}~\rm M_\odot}{0.08}=\frac{DL}{10}$, we can observe the phenomenon of three magnification peaks, see Fig. \ref{fig:9}, at distance $D_L=7.4\times 10^{8}~\rm km$ (The Schwarzschild radius approximately equals to $7.8\times 10^6~\rm km$). The distance from Earth to the center of the Milky Way galaxy is about $2.4\times 10^{16}~\rm km$, and the magnification effect observed at this distance is equivalent to a small mass situation in Fig. \ref{fig:6}, which converge to a single peak situation. (Not even observable, according to Fig. \ref{fig:4}.). 

Our research provides a phenomenological difference in magnification between black holes and wormholes, and it provides a theoretical basis for further research in the magnification of wormholes and black holes.

\end{document}